\documentclass[conference]{IEEEtran}
\IEEEoverridecommandlockouts

\usepackage[dvips]{graphicx}
\usepackage{latexsym}
\usepackage{epsfig}
\usepackage{color}
\usepackage{amsmath} 
\usepackage{amssymb}
\usepackage{amsxtra}
\usepackage{enumitem}
\usepackage{float}
\usepackage[T1]{fontenc} 
\usepackage{setspace}
\usepackage{comment}
\usepackage[cmintegrals]{newtxmath}
\usepackage{graphicx}
\graphicspath{{Figures/}}
\usepackage{url}
\usepackage{cite}
\usepackage{algorithm}
\usepackage{algpseudocode}
\makeatletter
\def\BState{\State\hskip-\ALG@thistlm}
\makeatother
\usepackage{epstopdf}
\usepackage{amsmath}
\usepackage{amsxtra}
\usepackage{amstext}
\usepackage{amssymb}
\usepackage{latexsym}
\usepackage{dsfont} 
\usepackage{color}
\usepackage{multirow}
\usepackage{float}
\usepackage{hyperref}
\usepackage{tabularx}
\usepackage[table]{xcolor}
\usepackage{lscape}
\usepackage{diagbox}
\usepackage{mwe}
\usepackage{units}
\usepackage{cases}
\usepackage{mathtools}

\usepackage[subrefformat=parens,labelformat=parens,caption=false,font=footnotesize]{subfig}
\newcommand{\Hpow}{{\sf H}}
\newcommand{\Tpow}{{\sf T}}
\newcommand{\Invpow}{{\sf -1}}

\newcommand{\mbf}[1]{\mathbf{#1}}

\newcommand{\nth}[1]{{#1}{\text{th}}}
\newcommand{\abs}[1]{\left|{#1}\right|}
\newcommand{\norm}[1]{\left\|{#1}\right\|}
\DeclareMathOperator*{\argmax}{arg\,max}   

\newcommand{\ZF}{\mathrm{ZF}}

\newcommand{\grd}{\mathrm{GRAND}}

\makeatother

\makeatletter
\def\ps@IEEEtitlepagestyle{%
  \def\@oddfoot{\mycopyrightnotice}%
  \def\@oddhead{\hbox{}\@IEEEheaderstyle\leftmark\hfil\thepage}\relax
  \def\@evenhead{\@IEEEheaderstyle\thepage\hfil\leftmark\hbox{}}\relax
  \def\@evenfoot{}%
}

\def\mycopyrightnotice{%
  \begin{minipage}{\textwidth}
  \centering \scriptsize
    This work has been accepted by the 2024 IEEE 100th Vehicular Technology Conference (VTC2024-Fall) for publication. Copyright may be transferred without notice, after which this version may no longer be accessible.
  \end{minipage}
}
\makeatother

\begin{document}

\title{Leveraging parallelizability and channel structure in THz-band, Tbps channel-code decoding
}

\author{\IEEEauthorblockN{Hakim~Jemaa$^{\ast}$, Hadi~Sarieddeen$^{\ddagger}$, Simon Tarboush$^{\dag}$, Mohamed-Slim~Alouini$^{\ast}$, and~Tareq~Y.~Al-Naffouri$^{\ast}$}
\IEEEauthorblockA{$^{\ast}$\textit{King Abdullah University of Science and Technology (KAUST)}, Kingdom of Saudi Arabia, 23955-6900 \\
$^{\ddagger}$\textit{American University of Beirut (AUB)}, Beirut 1107 2020, Lebanon \\
$^{\dag}$ Independent researcher from Damascus, Syria \\
\{hakim.jemaa, slim.alouini, tareq.alnaffouri\}@kaust.edu.sa;  hadi.sarieddeen@aub.edu.lb; simon.w.tarboush@gmail.com}

\thanks{The work was supported by the KAUST Office of Sponsored Research under Award ORA-CRG2021-4695, and the AUB University Research Board.}
}

\maketitle

\begin{abstract}

As advancements close the gap between current device capabilities and the requirements for terahertz (THz)-band communications, the demand for terabit-per-second (Tbps) circuits is on the rise. This paper addresses the challenge of achieving Tbps data rates in THz-band communications by focusing on the baseband computation bottleneck. We propose leveraging parallel processing and pseudo-soft information (PSI) across multicarrier THz channels for efficient channel code decoding. We map bits to transmission resources using shorter code-words to enhance parallelizability and reduce complexity. Additionally, we integrate channel state information into PSI to alleviate the processing overhead of soft decoding. Results demonstrate that PSI-aided decoding of 64-bit code-words halves the complexity of 128-bit hard decoding under comparable effective rates, while introducing a $\unit[4]{dB}$ gain at a $10^{-3}$ block error rate. The proposed scheme approximates soft decoding with significant complexity reduction at a graceful performance cost.
\end{abstract}

\section{Introduction}
\label{sec:introduction}

\IEEEPARstart{T}{erahertz} (THz)-band communications, spanning frequencies from 0.1 THz to 10 THz, hold immense promise for future wireless networks, offering data rates in the terabit-per-second (Tbps) range that notably exceed those of traditional millimeter-wave (mmWave) and sub-6 gigahertz (GHz) systems \cite{sarieddeen2021overview}. Although overcoming the so-called THz gap, characterized by device-level challenges, remains crucial for unlocking groundbreaking technological advancements \cite{akyildiz2022terahertz}, this work focuses on the spectral and hardware constraints.

At the spectrum level, advancements in THz channel modeling are propelled by extensive measurement campaigns \cite{sheikh2022thz}. THz channels exhibit extreme sparsity in time and angle domains, surpassing mmWave frequencies \cite{9411894}. In addition, severe path loss due to spreading and molecular absorption, which increases with distance and gas density, results in frequency-selective behavior even in line-of-sight (LoS) conditions \cite{sarieddeen2021overview, 6998944}. Indoor non-line-of-sight (NLoS) THz communications remain feasible but less common with massive arrays and high-gain antennas \cite{sheikh2022thz, tarboush2021teramimo}. At the hardware level, realizing the full potential of THz-band communications necessitates aligning transceiver operations with hardware limitations in data conversion sampling frequencies and digital circuit clock frequencies \cite{epic2019}. In particular, attaining Tbps data rates with current technologies requires processing thousands of information bits per clock cycle, which can only be achieved through substantial hardware parallelism \cite{sarieddeen2023bridging}. 


Channel-code decoding poses a significant computational bottleneck in baseband processing, where factors such as complexity, energy consumption, and parallelizability greatly influence system design efficiency and performance \cite{10274838}. Turbo, low-density parity-check (LDPC), and polar codes are prominent contenders for future communication standards, influencing system complexity and performance~\cite{balatsoukas2016hardware}. Furthermore, the evolving landscape of applications drives a shift towards universal decoding that accommodates various code rates and lengths. For short-code and high-rate scenarios, noise-centric decoders such as guessing random additive noise decoding (GRAND) have demonstrated cost-effectiveness~\cite {duffy2019capacity}.

Recent works \cite{epic2019} aimed to achieve practical wireless Tbps communications in various THz-band applications by focusing on advanced channel coding and emphasizing decoder enhancements and parallelization to establish key performance indicators (KPIs). Such works asserted that addressing THz-specific challenges and implementing extensive parallelism across all baseband blocks are crucial \cite{sarieddeen2023bridging,JemaaDetection2022}. However, the effect of THz propagation environments has not been factored into channel coding and decoding designs in these works.


An accurate understanding of channel models enables the generation of reliable soft information \cite{605601}, which can enhance THz-specific decoding schemes. However, acquiring precise soft information in log-likelihood ratios (LLRs) is challenging under Tbps constraints. Additionally, feeding high-resolution LLRs to the channel decoder for each signal reception can create a bandwidth bottleneck between baseband modules, with LLR storage posing further challenges \cite{sarieddeen2023bridging}.

In this work, we propose incorporating channel state information (CSI) and additive noise statistics into channel bit mapping and channel-code decoding in a single-shot data detection and decoding mechanism. Specifically, we utilize pseudo-soft information (PSI) and short codes to balance complexity and performance. We show that PSI, representing the effective signal-to-noise ratio (SNR) following data detection \cite{sarieddeen2022grand} and dependent solely on the THz-band channel and noise characteristics, can enhance the efficiency of channel-code decoding. Short codes are proposed as a necessary means for parallelizability, reducing latency and overall system complexity. We study the performance of the proposed design under a multicarrier (wideband) THz system.

Regarding notations, scalars $(a, A)$, vectors $(\mathbf{a})$, and matrices $(\mathbf{A})$ are represented by non-bold, bold lowercase, and bold uppercase letters, respectively. 
Superscripts denote the transpose $({}^\Tpow)$, Hermitian $({}^\Hpow)$, and inverse $({}^\Invpow)$ operators. The $\abs{\cdot}$ operator denotes the absolute value or matrix determinant and $\norm{\cdot}$ denotes the Euclidean norm. $\mathrm{Pr}(\cdot)$ represents probability and $\Gamma(\cdot)$ describes the Gamma function. The subscripts $t$ and $r$ specify transmitter and receiver parameters, respectively.


\section{System and channel models}

\subsection{System model}

We consider a single-input single-output (SISO) multicarrier system of bandwidth, $B$, and number of subcarriers, $L$. We denote by $\mathrm{h}_l \!\in\! \mathbb{C}$ the complex channel gain of the $l$th subcarrier, $l\!\in\!\{1,2,\cdots,L\}$. The frequency-domain complex baseband input-output relation at the $l$th subcarrier is given by~\cite{tarboush2022single}
\begin{equation}
    \mathrm{y}_l =  \sqrt{P_t G_{r} G_{t}}\mathrm{h}_l \mathrm{x}_l +  \mathrm{n}_l,
    \label{eq:sys_model}
\end{equation}
where the received and transmitted symbol vectors are constructed as $\mathbf{y} \!=\! [y_1, y_2, \ldots, y_L]^\Tpow \!\in\! \mathbb{C}^{L \times 1}$ and $\mathbf{x} \!=\! [x_1, x_2, \ldots, x_L]^\Tpow \!\in\! \mathbb{C}^{L \times 1}$, respectively. Each $x_l$ is derived from a quadrature amplitude modulation (QAM) constellation, $\mathcal{X}_l$. The vector $\mathbf{n} = [n_1, n_2, \ldots, n_L]^\Tpow \in \mathbb{C}^{L \times 1}$ similarly represents additive white Gaussian noise (AWGN) over received symbols, with a noise power of $\sigma^2 = \frac{N_0}{2}$, where $N_0$ is the one-sided noise power spectral density. Furthermore, the terms $P_t$, $G_{t}$, and $G_{r}$ correspond to the transmitter power, transmitter antenna gain, and receiver antenna gain, respectively. We assume perfect CSI at the receiver. 

The bit-representation of $x_l$ is $\mbf{c}_{l}\!=\![c_{l,1}\cdots c_{l,j}\cdots c_{l,q_{l}}]^{T}\!\in\! \{0,1\}^{q_{l}}$, where $q_{l}\!=\!\lceil{\log_2(\abs{\mathcal{X}_l})\rceil}$. The bit-representation of $\mbf{x}$ is $\mbf{c}\!=\![\mbf{c}_{1}\cdots \mbf{c}_{l} \cdots \mbf{c}_{L}]^{T}\!\in\!\{0,1\}^N$, where $N\!=\! \sum_{l=1}^{L} q_{l}$. We denote by $\mbf{c}$ a code-word encoded with an error correcting code of code-rate $R\!=\!K/N$. A code-book $\mathcal{C}$ comprises all possible code-words. One or more code-words are mapped to the symbols of a multicarrier frame. At the receiver side, a hard detector equalizes the channel and recovers a symbol vector, $\hat{\mbf{x}}$, from $\mbf{y}$; a demapper recovers a word, $\hat{\mbf{c}}$, from $\hat{\mbf{x}}$; a channel-code decoder recovers $\mbf{c}$. A soft decoder accepts, in addition to $\hat{\mbf{c}}$, a vector of bit-reliability information, $\mbf{\Lambda}\!=\![\lambda_{1,1}\cdots \lambda_{l,j}\cdots \lambda_{L,q}^{}]^{T}\!\in\!\mathbb{C}^N$. The remainder of this paper puts this system model in a THz-band, Tbps context.

\subsection{THz channel model}

The performance of detectors and decoders in THz communication systems is heavily influenced by the underlying channel and noise distributions. Capturing THz-band channel characteristics analytically has been challenging due to the complexity and poor fit of conventional distributions to THz channel measurements. Recent THz channel modeling can be categorized into two approaches. The first is statistical modeling, utilizing a modified Saleh-Valenzuela (SV) channel model, the frequency domain channel coefficients of which are formulated in~\cite{tarboush2021teramimo}. The second approach involves distribution-based THz channels. Here, we decompose $\mathrm{h}_l$ into $\mathrm{h}_l = h^{\mathrm{p}} h^{\mathrm{f}}$ \cite{8610080}, where $h^{\mathrm{p}}$ accounts for the free-space path loss that depends on the communication distance, molecular absorption, and subcarrier frequency \cite{tarboush2021teramimo}, and $h^{\mathrm{f}}$ represents the small-scale fading effects \cite{8610080}. Further details can be found in \cite{tarboush2021teramimo}.

Recent studies~\cite{papasotiriou2023outdoor,5090420} identified the mixture gamma (MG) distributions as candidate fitting distributions for small-scale fading in outdoor THz communications. A MG distribution is characterized by a probability density function (PDF),
\begin{equation}
\label{MGdist}
    f_{Z}(t)= \sum_{i=1}^M \alpha _i t^{\beta _i -1} e^{-\zeta _i t},  t\ge 0,
\end{equation}
where $M$ is the number of MG components and $\alpha_i=w_i\zeta_i^{\beta_i}/\Gamma\left(\beta_i\right)$, with $w_i$ representing the weight, $\zeta _i$ the scale, and $\beta _i$ the shape of the $\nth{i}$ component. The weights satisfy the condition $\sum_{i=1}^M w_i \!=\! 1$. Furthermore, the $\alpha-\mu$ distributions best characterize small-scale fading in indoor THz environments \cite{papasotiriou2021experimentally,5090420}. The corresponding PDF is
\begin{equation}
f_{Z}(t)=\frac{\alpha \mu^\mu t^{\alpha \mu-1}}{|\hat{Z}|^{\alpha \mu} \Gamma(\mu)} e^{- \left(\frac{t}{|\hat{Z}|}\right)^\alpha},
\end{equation}
with $\alpha > 0$ indicating the fading parameter, and $\mu$ and $|\hat{Z}|$ denoting the shape and scale parameters, respectively.


\section{Soft and pseudo-soft information}
\label{sec:psi}

Optimal maximum-likelihood (ML) detection minimizes the Euclidean distance between the received and estimated signals, recovering, through an exhaustive search, a hard-detection vector, 
\begin{equation}
\hat{\mathbf{x}}^{\mathrm{ML}} = \underset{\mathbf{x} \in \mathcal{X}^{\mathrm{L}}}{\arg \min }\norm{\mathbf{y}-\mathbf{hx}}^{2}.
\end{equation}
For enhanced performance, soft-output a posteriori bit LLRs are computed as reliability information per bit. We define the LLR, $\lambda_{l,j}$, of the $\nth{j}$ bit of symbol $x_{l}$ as
\begin{equation}
 \lambda_{l,j} \!\triangleq\! \lambda\left(c_{l,j}\right|\mathbf{y}) \!\triangleq\! \ln \left({\mathrm{Pr}\left[c_{l,j}\!=\!1 | \mathbf{y}, \mathbf{h}\right]}/{\mathrm{Pr}\left[c_{l,j}\!=\!0 | \mathbf{y}, \mathbf{h}\right]}\right),
\end{equation}
where $c_{l,j}\!=\!0$ and $c_{l,j}\!=\!1$ are assumed equally likely. For a SISO link per subcarrier (assuming independent subcarriers), ML soft information can be extracted by searching disjoint symbol constellations to compute
\begin{equation}
\lambda_{l, j}^{\mathrm{ML}}=\frac{1}{\sigma^2}\left(\min _{x_l \in \mathcal{X}_l^{j, 1}}\left|y_l-h_l x_l\right|^2-\min _{x_l \in \mathcal{X}_l^{j, 0}}\left|y_l-h_l x_l\right|^2\right),
\end{equation}
where $\mathcal{X}_l^{j, 1} \!=\! \left\{x_l \!\in\! \mathcal{X}_l: c_{l, j}\!=\!1\right\}$ and $\mathcal{X}_l^{j, 0} \!=\! \left\{x_l \!\in\! \mathcal{X}_l: c_{l, j}\!=\!0\right\}$.

Under the THz-band, Tbps constraints \cite{sarieddeen2023bridging}, simpler linear schemes are often necessary. A typical example is the ZF detector, which equalizes the received signal into $\hat{\mathbf{y}}^{\mathrm{ZF}}=\left(\mathbf{h}^\Hpow \mathbf{h}\right)^{-1} \mathbf{h}^\Hpow \mathbf{y}$, and then computes LLRs as
\begin{equation}
\lambda_{l, j}^{\mathrm{ZF}} =\frac{1}{{\sigma_{l}^{\ZF}}^2}\left(\min _{x_l \in \mathcal{X}_l^{j, 1}}\left|\hat{y}_l^{\mathrm{ZF}}-x_l\right|^2-\min _{x_l \in \mathcal{X}_l^{j, 0}}\left|\hat{y}_l^{\mathrm{ZF}}-x_l\right|^2\right).
\label{eq:ZF_soft}
\end{equation}

Although generating soft detection outputs enhances decoding, there is a preference for generating solely hard detection outputs, as search routines can be computationally demanding and have varying complexity costs. Additionally, quantizing the $\lambda_{l,j}$ values, say to a five-bit resolution, and transmitting them to the decoder consumes significantly more bandwidth than processing single-bit hard outputs. As a solution, in addition to $\hat{\mbf{c}}$, we propose generating marginal PSI, $\bar{\mbf{\Lambda}}$, which does not require computation at every channel use \cite{sarieddeen2022grand}.

Following linear detection, the scaled effective noise variance, ${\sigma_{l}^{\ZF}}^2$, incorporates channel and noise information that can be leveraged for soft information approximation. In particular, per-bit ZF-based PSI can be expressed as
\begin{equation}
\Bar{\lambda}_{l,j}^{\mathrm{ZF}} =\frac{1}{{\sigma_{l}^{\ZF}}^2}, \quad \text{where} \quad {\sigma_{l}^{\ZF}}^2=\sigma_l^2\left(h_l^{\Hpow} h_l\right)^\Invpow,
\label{eq:ZF_psi}
\end{equation}
which is common for all bits of the same symbol, and can be retained over a coherence time or bandwidth. Thus, leveraging PSI reduces complexity and processing overhead and minimizes performance degradation compared to soft decoding.

\section{Parallelizability and channel-code decoding}

Channel-code decoding is a critical bottleneck for efficient communication systems design. Indeed, the overall system complexity, performance, and energy efficiency are determined by the coding/decoding scheme architecture \cite{epic2019}.

\subsection{Tbps candidate coding and decoding schemes}

The state-of-the-art consensus has highlighted turbo, LDPC, and polar codes as candidates for future generations of wireless communications. Various projects have established KPIs to assess the feasibility of achieving Tbps data rates under such schemes \cite{epic2019,kestel2020506gbit}. Critical KPIs include energy and power efficiency, throughput, complexity, and degree of parallelizability. Polar codes, in particular, have garnered rapid interest due to their capacity-achieving performance and scalability, especially for lengthy codes \cite{Sural8880815}. Their performance is notably appealing for short to moderate code lengths, thus enhancing scalability and parallelizability. Polar code decoding employs successive cancellation (SC) and successive cancellation list (SCL) decoding techniques. Although SC offers low complexity, its error-correction capability is limited. SCL, as an extension of SC, utilizes list decoding to significantly boost error correction at the expense of increased complexity, which is dependent on the list size.

In the context of short codes and high rates, which present a fitting scenario for achieving high parallelizability and low complexity, noise-centric and universal decoders like GRAND are garnering increased attention \cite{sarieddeen2022grand}. The GRAND algorithm operates by searching for the most likely noise effect sequence that could have corrupted $\mathbf{c}$, denoted as $\mathbf{w} = [w_{1,1} \ldots w_{l,j} \ldots w_{L,q_{l}}]^T \in \{0,1\}^N$, in non-increasing probability. Noting that $\Pr(\hat{\mathbf{c}}|\mathbf{c}) = \Pr(\hat{\mathbf{c}} = \mathbf{c} \oplus \mathbf{w})$, GRAND recovers the code-word, \cite{sarieddeen2022grand}
\begin{equation}
    \mathbf{c}^{\grd} = \argmax \ \{\Pr(\mathbf{w} = \hat{\mathbf{c}} \ominus \mathbf{c}): \mbf{c}\in\mathcal{C} \},
\end{equation}
where $\oplus$ and $\ominus$ represent bit noise effect onset and inversion, both as XOR operators. GRAND is well-suited for THz communication applications due to two primary factors. Firstly, it excels with short, moderately redundant codes commonly found in parallelizable baseband structures. Secondly, GRAND operates effectively without requiring interleaving and performs best when fed distinct structures of channel and noise, facilitating the rapid guessing of correct noise effects. These channel structures are effectively conveyed in PSI. 

\subsection{Parallelism via short codes}
The architecture of the coding scheme plays a crucial role in decoding parallelism, with inherently parallel decoders being preferred in this context. However, strategies for parallel decoding can significantly impact throughput, latency, and energy efficiency. Significant advancements have been made in THz/Tbps coding schemes under $\unit[28]{nm}$ technology. LDPC codes have achieved an energy efficiency of $\unit[6]{pJ/bit}$ and a throughput of $\unit[160]{Gbps}$. Polar codes have reached a throughput of $\unit[620]{Gbps}$ with an energy efficiency of $\unit[4.6]{pJ/bit}$. However, turbo codes only achieved a throughput of $\unit[48]{Gbps}$ and an area efficiency of $\unit[5.33]{W/mm^2}$ \cite{epic2019}. Soft-detection ordered reliability bits GRAND (ORBGRAND) has also demonstrated an energy efficiency of $\unit[0.76]{pJ/bit}$ and power consumption of $\unit[4.9]{mW}$ at $10^{-7}$ frame-error-rate \cite{10067519}, achieving an area efficiency of $\unit[16.3]{Gbps/mm^2}$ in $\unit[40]{nm}$ CMOS.

\begin{algorithm}[t]
\caption{Proposed parallelizable transceiver design}\label{alg:algo}
\begin{algorithmic}[1]
\State \textbf{Inputs:} $\mathbf{y}$, $\mathbf{h}$, $L$, $V$, $\mathcal{X}$
\State \textbf{Outputs:} $\mathbf{{c}}$
\State \textbf{Detection:}
\For{$l=1,\cdots,L$}  \Comment{parallelism}
    \State $v \gets 1+(\sum_{i=1}^{l}q_i)/ N$ \Comment{integer division}
    \State $\hat{\mathbf{x}}_{v}^{\mathrm{ZF}} [l]\gets \left\lfloor\left(h_{l}^\Hpow h_{l}\right)^{-1} h_{l}^\Hpow y_{l}\right\rceil_{\mathcal{X}_{l}}$

    \State $\Bar{\boldsymbol{\lambda}}_{v}^{\mathrm{ZF}}[l,j] \gets \left(1/{\sigma_l^2\left(h_l^{\Hpow} h_l\right)^\Invpow}\right)$ for $j \in \![1\cdots q_{l}]$
    
\EndFor
\State \textbf{Decoding:} 
\For{$v=1,\cdots,V$} \Comment{parallelized}

    \State $\mathbf{c}_{v}^{\mathrm{GRAND}/\mathrm{SCL}} \gets \mathrm{DEC}_{v} (\Bar{\boldsymbol{\lambda}}_{v}^{\mathrm{ZF}},\hat{\mathbf{x}}_{v}^{\mathrm{ZF}})$

\EndFor
\end{algorithmic}
\end{algorithm}

\begin{figure*}[t]
\centering
    \subfloat[Indoor $\alpha-\mu$ sub-THz channel.]{\label{fig:alpha_thz}\includegraphics[width=0.49\linewidth]{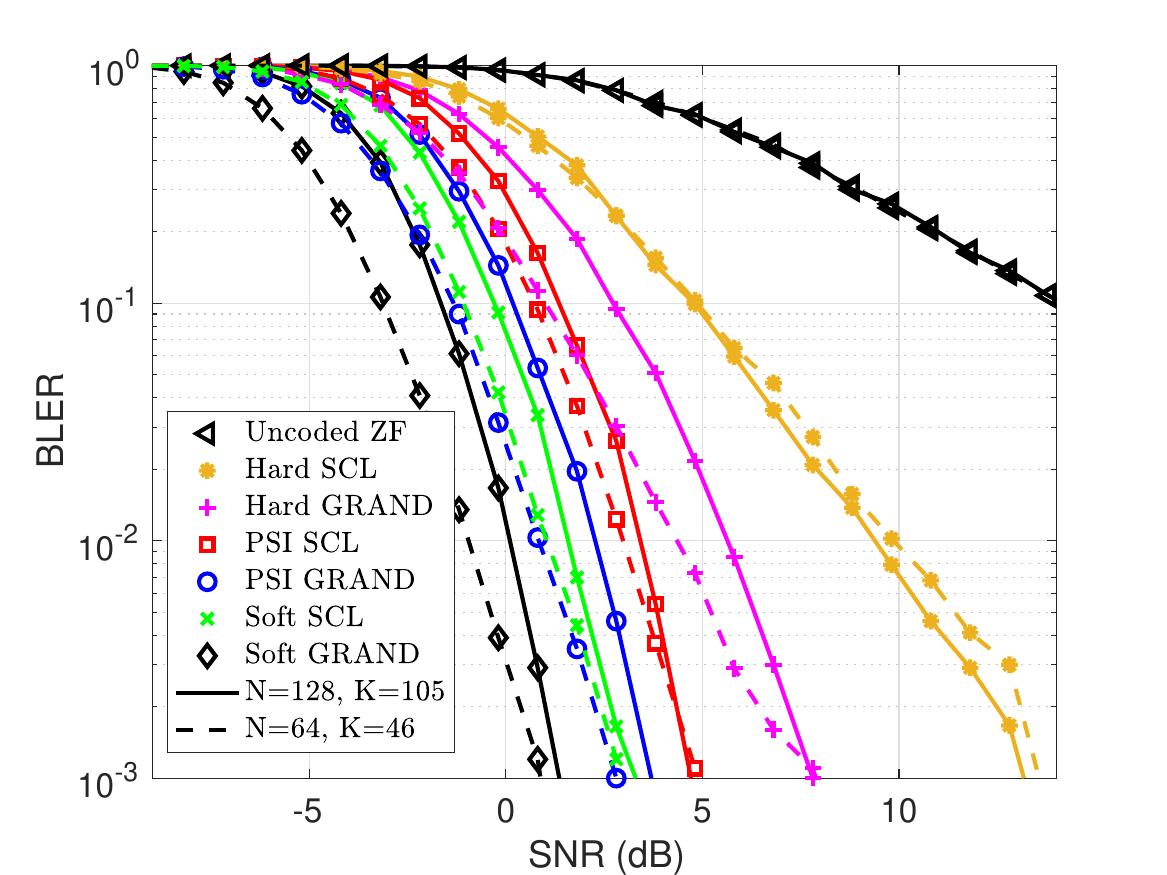}}
    \hfill
    \subfloat[Outdoor MG sub-THz channel.]    {\label{fig:MG_thz}\includegraphics[width=0.49\linewidth]{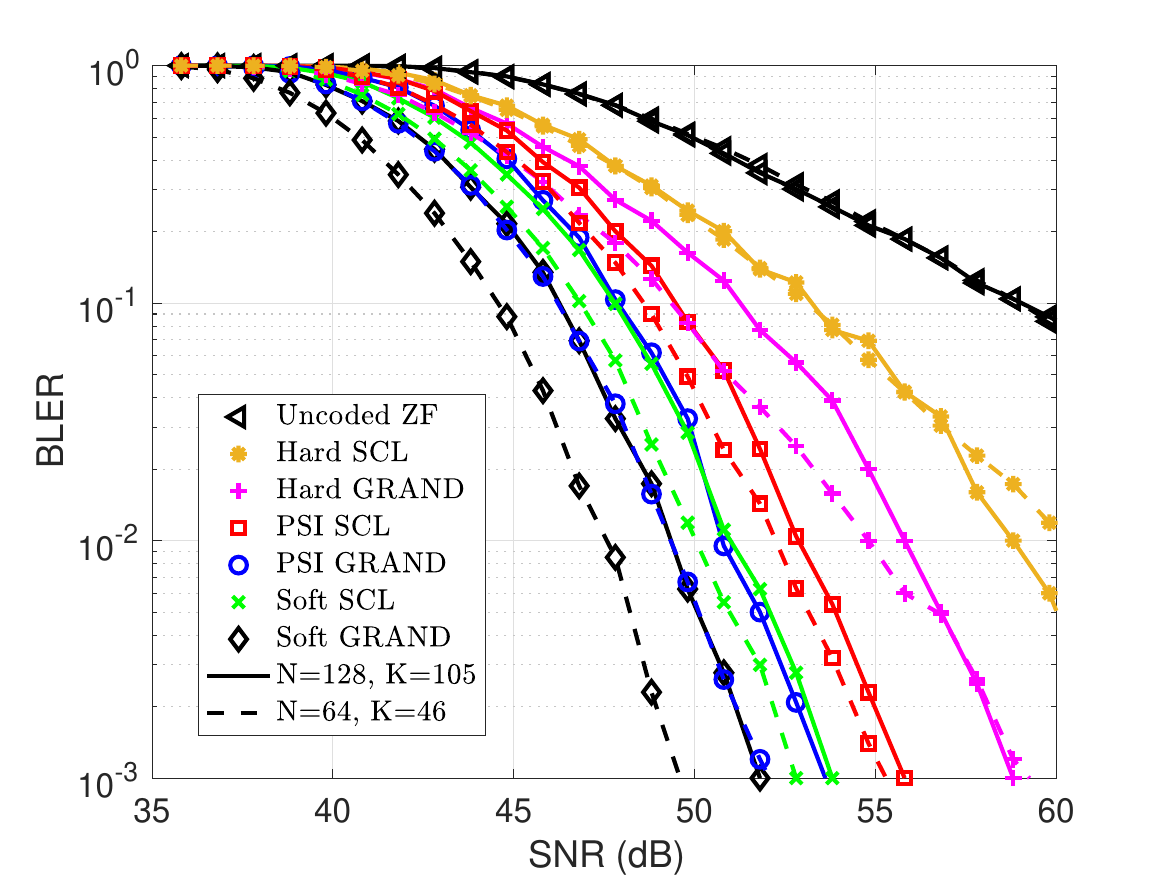}}
    \hfill
\caption{\small{BLER versus SNR for hard, PSI-based, and soft decoding - ZF detection - BPSK modulation - SCL ($L=16$) and GRAND decoding of polar codes ($N=128$, $K=116$)/($N=64$, $K=57$) (solid/dotted lines) - frequency-selective THz channels.}}
\label{fig:main}
\end{figure*}
\begin{figure}[t]
  \centering
  \includegraphics[width=0.49\textwidth]{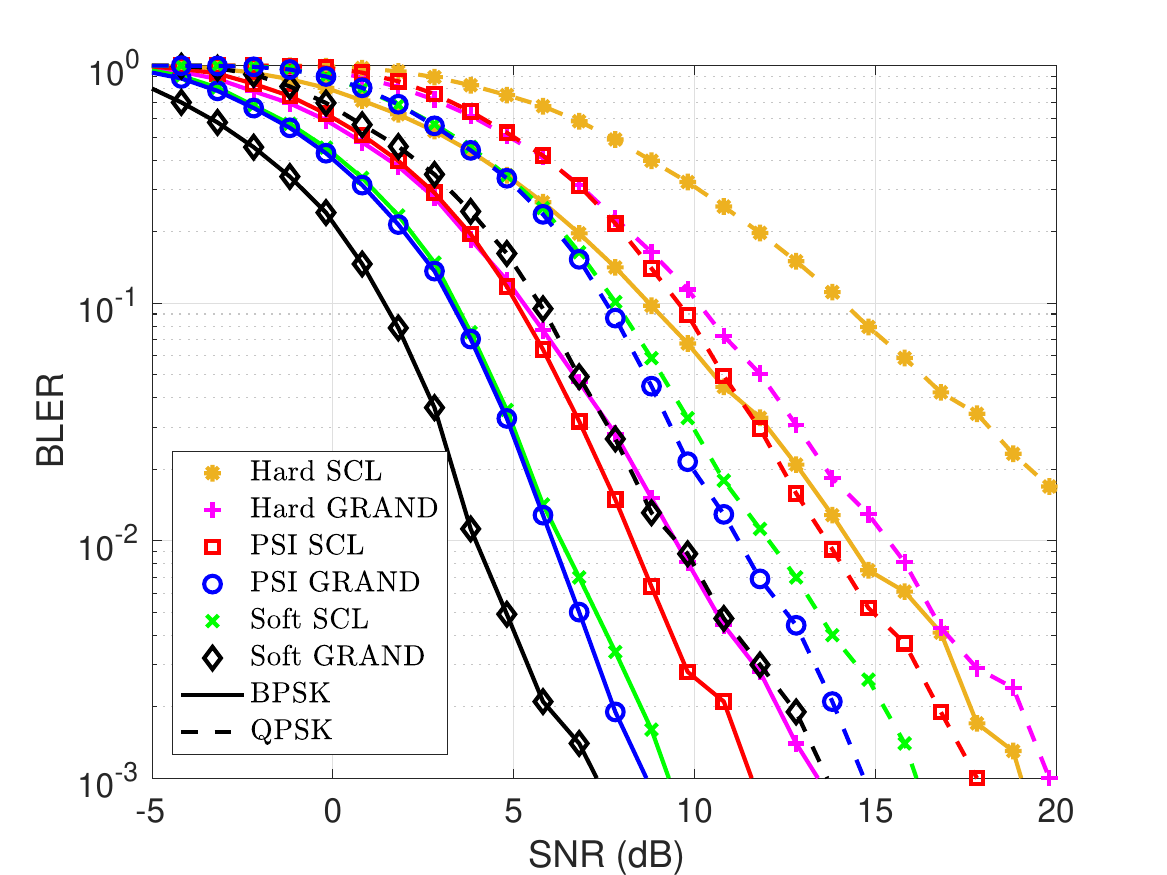}
 \caption{\small{Hard, PSI-based, and soft decoding - ZF detection - SCL ($L\!=\!16$) and GRAND decoding - polar codes ($N\!=\!64$, $K\!=\!57$) - BPSK and QPSK - frequency-selective indoor SV THz channel \cite{tarboush2021teramimo}.}}
  \label{fig:MP_thz}
  \vspace{-5mm}
\end{figure}

In this work, we propose a parallelizable decoder design that accounts for the source mapping of bits to the subcarriers using short code-words. Short codes not only reduce complexity and latency but also improve power efficiency \cite{dizdar2016high}. The complexity of candidate decoders scales linearly with code lengths: decoding a code-word of length $N$ is $V$ times more complex than decoding one of length $N/V$, as evidenced by cell count and area metrics in polar-code ASIC implementations, for example \cite{dizdar2016high}. Comparisons reveal that larger polar codes ($N=1024$, $K=518$) achieve a throughput of $\unit[506]{Gbps}$, a latency of $\unit[309.7]{ns}$, and a power consumption of $\unit[4.74]{W}$. In contrast, smaller codes ($N=128$, $K=70$) achieve a throughput of $\unit[64]{Gbps}$, a latency of $\unit[41.8]{ns}$, and a power consumption of $\unit[0.25]{W}$, highlighting significant reductions (by a factor of $V=10$) in latency and power usage \cite{kestel2020506gbit, sarieddeen2023bridging}.

In most cases, however, shorter code-words lead to performance degradation. Reducing the code-word size exacerbates diversity loss, inherently degrading performance, especially when error correction applies per sub-channel rather than across the entire channel. The goal is not to indiscriminately shorten code lengths but to carefully balance performance with overall complexity, latency, and power consumption.

Algorithm \ref{alg:algo} briefly summarizes the proposed framework, which can be extended to account for different detection and decoding schemes. The first step is to determine $V$, the parallelizability factor, equal to the number of code-words per frame. $V$ depends on the number of subcarriers, $L$, the QAM modulation, $\mathcal{X}_{l}$, and the encoder code length, $\mathrm{N}$. We instantiate $V$ encoders and map source bits to each encoder for $v\in\{1,2,\cdots,V\}$. The modulated bits are transmitted over $\mathbf{h}$ following \eqref{eq:sys_model}. Upon receiving $\mathbf{y}$, the detector computes $\mathbf{\hat{x}} = \left\lfloor\mathbf{\hat{y}^{\mathrm{ZF}}}\right\rceil_{\mathcal{X}}$, where $\left\lfloor.\right\rceil_{\mathcal{X}}$ is a slicing operator over $\mathcal{X}$;
it also computes the PSI, $\Bar{\boldsymbol{\lambda}}$, according to \eqref{eq:ZF_psi} (vectors are accumulated with appropriate indices). Finally, each decoder, $\mathrm{Dec}_{v}$, will decode the received reliability information and process the bits in a parallel fashion. Although using shorter code-words introduces some performance degradation, we argue through simulations that leveraging PSI and noise-centric decoding methods can mitigate this loss in performance. 

\section{Simulation results}
\label{sec:simulation}

We employ the TeraMIMO simulator \cite{tarboush2021teramimo} to precisely model the characteristics of various THz channels, adapting it to incorporate small-scale fading modeled by the SV model, $\alpha$-$\mu$ distribution, and MG distributions. Simulations are conducted using BPSK and QPSK modulations, and the relevant parameters are detailed in Table \ref{tab:sim_params}. The noise power is defined as $N_0 = k_{B} T B$, where $k_{B}$ represents the Boltzmann constant, $B$ denotes the bandwidth, and $T = 300$ K is the system temperature in Kelvin. We adjust the transmitter power $P_{t}$ to explore different SNR ranges.

\begin{table}[t]
\centering
    \small
    \caption{Simulation parameters}
\begin{tabular}{|c|c|c|c|}
\hline
\diagbox{Parameters}{Channels}          & \begin{tabular}[c]{@{}c@{}}Indoor\\ $\alpha-\mu$\end{tabular} & \begin{tabular}[c]{@{}c@{}}Outdoor\\ MG\end{tabular} & \begin{tabular}[c]{@{}c@{}}Indoor\\ SV\end{tabular} \\ \hline
Frequency (THz)     & \multicolumn{2}{c|}{0.142 \cite{papasotiriou2021experimentally,papasotiriou2023outdoor}} & \begin{tabular}[c]{@{}c@{}}0.3 \cite{tarboush2021teramimo} \end{tabular} \\ \hline
Distance (m)        & 10.1 \cite{papasotiriou2021experimentally}      & 64 \cite{papasotiriou2023outdoor} & 2 \cite{tarboush2021teramimo}      \\ \hline
\begin{tabular}[c]{@{}c@{}}Number of \\ NLoS paths\end{tabular}       & 3         & 2         & \begin{tabular}[c]{@{}c@{}}Poisson\\ distribution \end{tabular}         \\ \hline
Bandwidth (GHz)     & \multicolumn{3}{c|}{4 \cite{papasotiriou2021experimentally}}                     \\ \hline
Antenna gains (dBi) & \multicolumn{3}{c|}{$G_{r} = 19$, $G_{t} = 0$ \cite{papasotiriou2023outdoor}} \\ \hline
\end{tabular}
\label{tab:sim_params}
\vspace{-5mm}
\end{table}

Fig.~\ref{fig:alpha_thz} illustrates the performance within a frequency-selective THz channel under ZF detection, with small-scale fading modeled using an $\alpha$-$\mu$ distribution. Simulations compare SCL and soft GRAND (ORBGRAND) \cite{Duffy2022Ordered} decoding of two polar-code configurations with the following effective rates: $K/N = 116/128$ and $K/N = 57/64$, both with a cyclic redundancy check (CRC) of 11 bits. Notably, significant performance discrepancies exist among hard, PSI-based, and soft detection. The gap between hard and soft detection exceeds \unit[12]{dB} at a block error rate (BLER) of $10^{-3}$. However, the gap between PSI-based and soft detection is approximately \unit[4]{dB} at the same BLER level, substantiating the efficacy of PSI in designing low-complexity non-iterative THz detectors and decoders. Additionally, ORBGRAND exhibits superior performance, outperforming SCL decoding by \unit[2]{dB}. This superiority is attributed to GRAND's capability to consider both CRC and polar code redundancies simultaneously. Further gains in soft GRAND can be achieved by utilizing the optimal, though more complex, SGRAND \cite{Solomon2020Soft}.

The results under an outdoor THz channel, modeled with an MG distribution, are presented in Fig.~\ref{fig:MG_thz}. The SNR gap between outdoor and indoor scenarios is caused by differences in communication distances ($\unit[64]{m}$ for outdoor, $\unit[10]{m}$ for indoor). A performance gap of approximately \unit[4]{dB} at a BLER of $10^{-3}$ exists between code-length configurations $K/N \!=\! 116/128$ and $K/N \!=\! 57/64$ for both soft and PSI-based GRAND, with the shorter code length demonstrating superior performance. Implicit in these observations is the marginal code-rate reduction at shorter code lengths, which hurts spectral efficiency. However, with larger available bandwidths at high frequencies, complexity reduction takes precedence over spectral efficiency. Additionally, a \unit[1]{dB} gap exists between PSI-aided GRAND and soft SCL decoding, consistently observed across channel scenarios, reinforcing the potential for achieving Tbps data rates through the integration of short codes and PSI. This approach reduces complexity and enhances parallelizability, addressing latency and power consumption challenges.

Fig.~\ref{fig:MP_thz} illustrates the performance under the indoor SV channel model with BPSK and QPSK modulations. While the performance ordering remains consistent across all channel scenarios, it is observed that with ZF-PSI, the low-complexity GRAND matches the performance of SCL decoding (with a list size of 16). The gap between soft and PSI-aided decoding for $N=64$ remains similar for both QPSK and BPSK. For BPSK, the gap between soft SCL and PSI-aided SCL is less than \unit[2]{dB}. Despite the fact that with QPSK, pairs of bits would share the same PSI (lower PSI resolution) PSI-aided GRAND remains superior to soft and PSI-based SCL.

\section{Conclusions}
\label{sec:conclusion}

As the demand for Tbps-achieving circuits increases and the gap in THz-operating devices narrows, addressing the challenge of achieving high data rates in THz-band communications becomes paramount. This paper proposes a solution for the baseband computational bottleneck by leveraging parallel processing and PSI in multicarrier THz channels for efficient channel-code decoding. By mapping bits to transmission resources using shorter code-words, we enhanced parallelizability and reduced complexity. By integrating channel state information into PSI, we alleviated the processing overhead of soft decoding. Simulation results demonstrated that PSI-aided 64-bit decoding halved the complexity of 128-bit hard decoding while introducing 4 dB gains at a BLER of $10^{-3}$. The proposed design approximated soft decoding with a significant complexity reduction at a graceful performance cost.

\bibliographystyle{ieeetr}


\end{document}